\documentclass[epj]{svjour}

\usepackage{graphicx,epsfig}
\usepackage{fancyhdr}

\def\beq   {\begin{equation}}
\def\eeq   {\end{equation}}
\def\beqd  {\begin{displaymath}}
\def\eeqd  {\end{displaymath}}
\def\beqaa {\begin{eqnarray}}
\def\eeqaa {\end{eqnarray}}

\def\ti  {\tilde}

\newcommand{\be}[1]{\begin{equation} \label{(#1)}}
\newcommand{\ee}{\end{equation}}
\newcommand{\baq}[1]{\begin{eqnarray} \label{(#1)}}
\newcommand{\eaq}{\end{eqnarray}}
\newcommand{\rf}[1]{(\ref{(#1)})}
\newcommand{\ba}{\begin{array}}
\newcommand{\ea}{\end{array}}
\def\mytitle{My title} 
\def\myauthors{My name}  
\def\mytype{My type of session}
\def\mysession{My session}
\def\mytitle{LFV Effects  
on Chargino Production at the ILC} %Put your title here!
\def\myauthors{K. Hohenwarter-Sodek and T. Kernreiter}    %Put your name here!
\def\mytype{Contributed Talk}    
\def\mysession{Colliders - SUSY Phenomenology}
\begin{document}
\title{Lepton Flavour Violating Effects  
on Chargino Production at the ILC}
\author{Karl Hohenwarter-Sodek %\inst{1}
\thanks{\emph{Email:} khsodek@hephy.oeaw.ac.at}%
 \and
 Thomas Kernreiter %\inst{1}% etc
\thanks{\emph{Email:} tkern@hephy.oeaw.ac.at
}}                     % Do not remove
\institute{Faculty of Physics, University of Vienna,
Boltzmanngasse 5, A-1090 Wien, Austria}
\date{}
\abstract{
We review the influence of lepton flavour violation (LFV)
on the production processes $e^+e^-\to\ti\chi^+_i\ti\chi^-_j$
at the International Linear Collider (ILC) with longitudinal 
$e^+$ and $e^-$ beam polarizations in the framework of the Minimal Supersymmetric
Standard Model (MSSM).
The $t-$channel sneutrino exchange contribution to the 
processes $e^+e^-\to\ti\chi^+_i\ti\chi^-_j$ is modified in the case of LFV,
as the sneutrino mass eigenstates have 
no definite flavour, and therefore more than one sneutrino
can contribute.
This influence can alter the cross section
$\sigma(e^+e^-\to\ti\chi^+_1\ti\chi^-_1)$ by a factor 
of 2 or more when varying the LFV mixing angles, in 
accordance with the restrictions due to the current 
limits on rare lepton decays.
Hence, the inclusion of LFV parameters can be important
when deducing the underlying model parameters from measured observables
such as $\sigma(e^+e^-\to\ti\chi^+_1\ti\chi^-_1)$.
\PACS{{11.30.Pb}{Supersymmetry}\and
{14.80.Ly}{Supersymmetric partners of known particles}
} % end of PACS codes
} %end of abstract
\maketitle
\section{Introduction}\label{intro}
The MSSM includes
the spin--1/2 partners of the $W^\pm$ bosons and the charged
Higgs bosons $H^\pm$. These states mix and form the
charginos $\ti\chi^\pm_k$, $k=1,2$, as the mass eigenstates. 
The charginos are of particular interest, as they will
presumably be among the lightest supersymmetric (SUSY) particles. 
Therefore the study of chargino production
\be{eq:prodchar}
e^+e^-\to\ti\chi^+_i\ti\chi^-_j ~,\qquad i,j=1,2~,
\ee
will play an important role at the ILC.
This process has been studied extensively in the literature, see e.g. 
\cite{char1,char2,char3}.
Procedures have been developed \cite{char2} to determine the underlying
parameters $\tan\beta$, $M_2$ and $|\mu|$, including the cosine of the
phase of $\mu$, $\cos\phi_\mu$, through
a measurement of a set of suitable observables in the processes
\rf{eq:prodchar}.
These studies assume that individual lepton flavour is conserved, which 
means that only one sneutrino ($\tilde\nu_e$) contributes to the 
processes \rf{eq:prodchar} via $t-$channel exchange.

In \cite{HohenwarterSodek:2007az} we have dropped this assumption, and have 
studied the influence of LFV parameters on the production cross sections
$\sigma(e^+e^-\to\ti\chi^+_i\ti\chi^-_j)$. 
In general, the sizes of the SUSY LFV parameters are restricted 
as they give rise to LFV rare lepton decays at 1--loop level, which
have not been observed so far. The current experimental upper bounds
on the branching ratios of LFV muon decays are 
BR$(\mu^-\to e^-\gamma) < 1.2 \cdot 10^{-11}$,
BR$(\mu^- \to e^- e^+ e^-) < 1.0\cdot 10^{-12}$ and for
the rate of $\mu^-- e^-$ conversion the best 
limit so far is $R_{\mu e}< 7.0\cdot 10^{-13}$, with
$R_{\mu e}=\Gamma[\mu^-+N(Z,A)\to e^-+N(Z,A)]/
\Gamma[\mu^-+N(Z,A)\to \nu_\mu+N(Z-1,A)]$ \cite{PDG}.
The sensitivities on LFV tau decays, are
smaller but have been improved substantially during the last years, where 
the current limits are BR$(\tau^-\to e^-\gamma) < 1.1 \cdot
10^{-7}$, BR$(\tau^-\to\mu^-\gamma) < 6.8 \cdot 10^{-8}$,
BR$(\tau^-\to e^-e^+e^-) < 2.0 \cdot 10^{-7}$ and
BR$(\tau^-\to\mu^-\mu^+\mu^-) < 1.9 \cdot 10^{-7}$ \cite{PDG}.

We have demonstrated in \cite{HohenwarterSodek:2007az} that in spite of the
restrictions due to LFV rare lepton decays the 
production cross section
$e^+e^-\to\ti\chi^+_1\ti\chi^-_1$ can change by a factor of 2 and more in 
the presence of LFV.
This can be the case even if the present bounds on 
LFV rare lepton decays improve by three orders of magnitude.
If LFV effects of this size occur, then the minimal sets 
\cite{char2} of observables may not be 
sufficient to determine the parameters in the chargino sector and have to be 
extended appropriately.
\section{Sneutrino mixing\label{sec:1}}

%%%%%%%%%%%
%         %
% FIG. 1  %
%         %
%%%%%%%%%%%
\begin{figure*}[t]
\hspace{1cm}
\begin{minipage}[t]{3.5cm}
\begin{center}
{\setlength{\unitlength}{1cm}
\begin{picture}(-1,2.5)
\put(-2,-1.1){\includegraphics{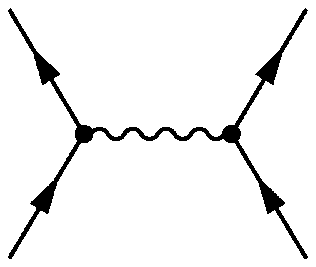}}
\put(-1.8,-1.6){$e^{-}$}
\put(1,-1.6){$\tilde{\chi}^{-}_j$}
\put(-1.8,1.5){$e^{+}$}
\put(1,1.5){$\tilde{\chi}^{+}_i$}
\put(-.3,.4){$\gamma$}
\end{picture}}
\end{center}
\end{minipage}
\hspace{2cm}
\vspace{.8cm}

\begin{minipage}[t]{3.5cm}
\begin{center}
{\setlength{\unitlength}{1cm}
\begin{picture}(-4.5,2.5)
\put(3.,2.3){\includegraphics{Feyn1.eps}}
\put(3.,1.8){$e^{-}$}
\put(6.,4.9){$\tilde{\chi}^{+}_i$}
\put(3.,4.8){$e^{+}$}
\put(6.,1.8){$\tilde{\chi}^{-}_j$}
\put(4.7,3.7){$Z$}
\end{picture}}
\end{center}
\end{minipage}
\hspace{2cm}
\vspace{.8cm}

\begin{minipage}[t]{3.5cm}
\begin{center}
{\setlength{\unitlength}{1cm}
\begin{picture}(1.3,2.5)
\put(11.8,5.4){\includegraphics{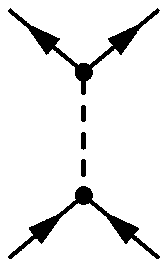}}
\put(11.5,8.2){$e^{+}$}
\put(11.5,5.2){$e^{-}$}
\put(13.3,8.2){$\tilde{\chi}^{+}_i$}
\put(13.3,5.1){$\tilde{\chi}^{-}_j$}
\put(12.1,6.6){$\tilde{\nu}_a$}
 \end{picture}}
\end{center}
\end{minipage}
\vspace{-4.5cm}
\caption{\label{bild1} 
Feynman diagrams for chargino production in $e^+e^-$-collisions.}
\end{figure*}

The sneutrino mass matrix in the MSSM including lepton flavour violation,
in the basis $(\tilde\nu_{e},\tilde\nu_{\mu},\tilde\nu_{\tau})$, 
is given by
\begin{eqnarray}
M^2_{\tilde \nu,\alpha\beta} &=&  M^2_{L,\alpha\beta} 
+ \frac{1}{2}~m^2_Z~\cos2\beta~\delta_{\alpha\beta}~.
\label{eq:sneutrinomass}
\end{eqnarray}
The indices $\alpha,\beta,\gamma=1,2,3$ characterize the flavours 
$e,\mu,\tau$, respectively.
$M^2_{L}$ is the hermitean soft SUSY breaking mass matrix for
the left sleptons, $m_Z$ is the mass of the $Z$ boson 
and $\tan\beta=v_2/v_1$ is the ratio of the vacuum expectation 
values of the Higgs fields.
The physical mass eigenstates are given by 
\begin{eqnarray}
\tilde \nu_i = R^{\tilde \nu}_{i\alpha}~\tilde\nu_\alpha' \qquad
(i=1,2,3)~,
\label{eq:Mixing}
\end{eqnarray}
with $\tilde \nu_\alpha'=(\tilde\nu_e, \tilde\nu_\mu, \tilde\nu_\tau)$.
The mixing matrix and the physical mass eigenvalues are obtained 
by an unitary transformation 
\be{eq:Diag} 
R^{\tilde\nu}\cdot M^2_{\tilde \nu}\cdot R^{\tilde\nu\dagger}=
{\rm diag}(m^2_{\tilde\nu_1},m^2_{\tilde\nu_2},m^2_{\tilde\nu_3})~,
\ee
where $m_{\tilde\nu_1} < m_{\tilde \nu_2} < m_{\tilde \nu_3}$.
Clearly, for $M_{L,\alpha\neq\beta}\neq 0$
the mass eigenstates, Eq.~(\ref{eq:Mixing}), are not flavour eigenstates.

The Feynman diagrams contributing to the processes \rf{eq:prodchar} 
are pictured in Fig.~\ref{bild1}. In the case of LFV the 
sneutrino contribution has to be modified, as now 
more than one sneutrino couples to the electron and positron 
(unless LFV arises solely due to the parameter $M^2_{L,23}$). 
This can be seen from the part of the interaction Lagrangian 
which gives rise to the $t-$channel sneutrino contribution \cite{mssm}:
\be{eq:lagsneuchar}
\mathcal{L}_{\ell\tilde{\nu}\tilde{\chi}^+}=
-g~V^*_{j1}~R^{\tilde\nu*}_{a1}~\overline{\tilde{\chi}^{-}_j}~
P_L~e~\ti\nu_a^\dagger
-g~V_{j1}~R^{\tilde\nu}_{a1}~\bar{e}~P_R~\tilde{\chi}^{-}_j
~\ti\nu_a,
\ee
where $P_{L,R}=1/2(1\mp\gamma_5)$, $g$ is the weak coupling constant 
and the unitary $2\times2$ mixing matrices $U$ and $V$ 
diagonalize the chargino mass matrix ${\mathcal M}_C$, 
$U^{\ast}{\mathcal M}_C V^{-1}=
{\rm diag}(m_{\chi_1},m_{\chi_2})$.

\section{Numerical analysis \label{sec:2}}

In the following we analyze numerically the influence of LFV
on the production cross section 
$\sigma(e^+e^-\to\tilde{\chi}^+_1\tilde{\chi}^-_1)$. 
The analysis is carried out for the ILC with a cms energy of $\sqrt{s}=500$~GeV,
and we assume that a degree of beam polarization 
of $-90\%$ for the electron beam and of $60\%$ for the
positron beam is feasible.

\subsection{$\tilde\nu_e$--$\tilde\nu_\tau$ mixing case}

\begin{figure}[t]
\setlength{\unitlength}{1mm}
\begin{center}
\begin{picture}(150,50)
\put(-45,-140){\mbox{\epsfig{figure=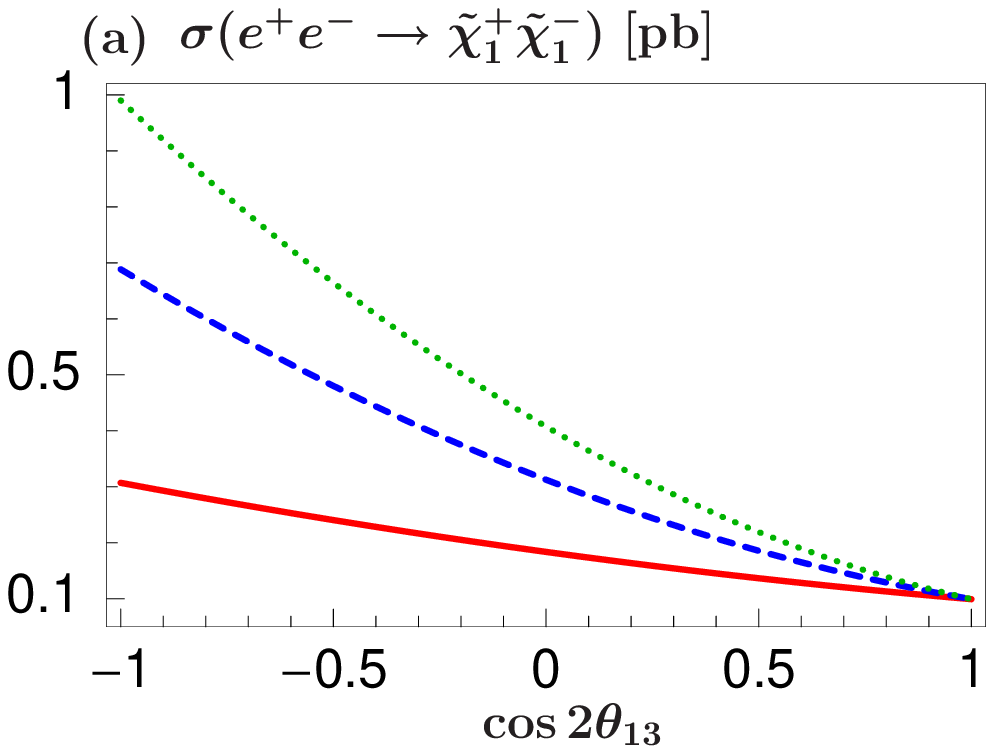,height=21.cm,width=15.4cm}}}
\put(-45,-200){\mbox{\epsfig{figure=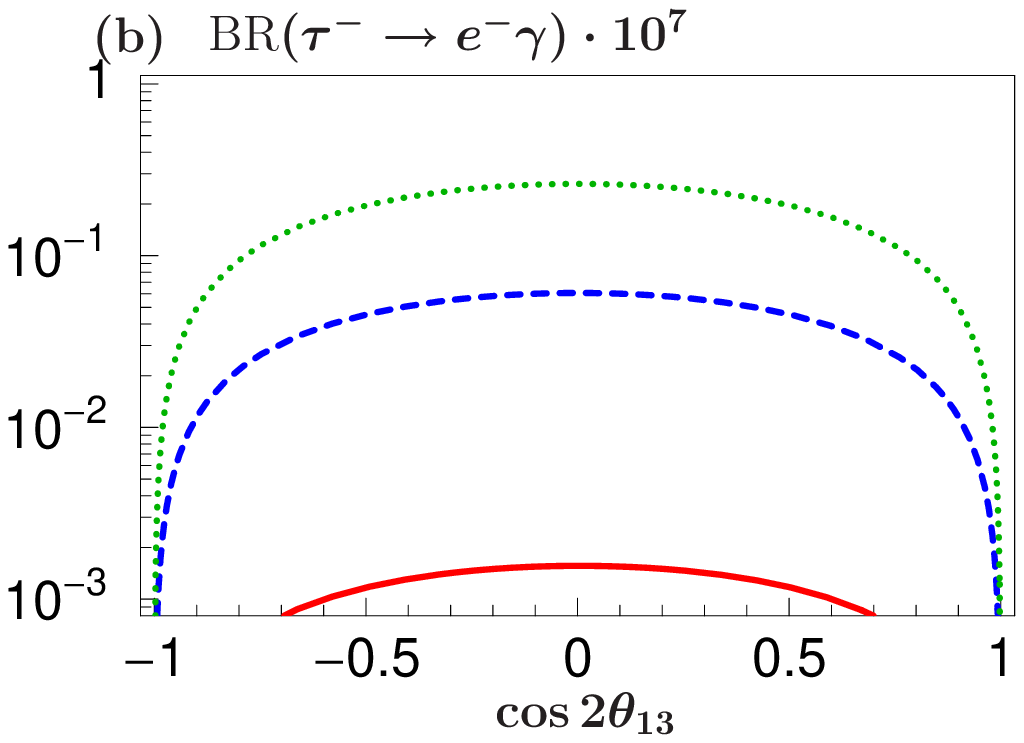,height=21.cm,width=15.4cm}}}
\end{picture}
\end{center}
\vskip6.5cm
\caption{(a) Cross section
$\sigma(e^+e^-\rightarrow\tilde{\chi}^+_1\tilde{\chi}^-_1)$ and
(b) branching ratio BR($\tau^-\to e^-\gamma$) as a function
of $\cos2\theta_{13}$. The three lines correspond to 
$m_{\tilde\nu_3}=400$~GeV (solid line),
600~GeV (dashed line) and 900~GeV (dotted line). 
The other parameters are as specified in the text.}
\label{fig:fig1}
\end{figure}
We start the discussion assuming LFV through non-vanishing $M^2_{L,13}$.
The size of $M^2_{L,13}$ is restricted by
the experimental upper bounds on the 
LFV processes $\tau^-\to e^-\gamma$ and $\tau^-\to e^- e^+ e^-$ to 
which it contributes at loop level. The formulae for the decay widths
of these reactions can be found in \cite{Hisano:1995cp}. For a complete 
1--loop calculation of the LFV leptonic three--body decays see \cite{Arganda:2005ji}.  
Furthermore, we require that the MSSM parameters have to 
respect the experimental limits of the anomalous magnetic
moments of the leptons, in particular that one of the muon, where
the difference between experiment and Standard Model (SM) prediction is 
$a^{\rm exp}_\mu-a^{\rm SM}_\mu=(29\pm 9)\cdot 10^{-10}$ \cite{Jegerlehner:2007xe}.
We impose that the SUSY contributions to $a_\mu$ 
must be positive and below $38 \cdot 10^{-10}$.

The MSSM parameters on which the cross section 
$\sigma(e^+e^-\rightarrow\tilde{\chi}^+_1\tilde{\chi}^-_1)$ depends are the
parameters in the chargino sector $\mu$, $M_2$ and $\tan\beta$,
and the soft SUSY breaking mass parameters in the sneutrino sector $M_{L,11}$, $M_{L,22}$,
$M_{L,33}$ and $M_{L,13}$ ($M_{L,12}=M_{L,23}=0$ in this subsection).
By employing the eigenvalue equations, Eq.~\rf{eq:Diag},
we treat the sneutrino masses $m_{\tilde\nu_1}$, $m_{\tilde\nu_2}$,
$m_{\tilde\nu_3}$ and the LFV mixing angle $\cos2\theta_{13}$ 
(with $\tan2\theta_{13}$$ =2M^2_{L,13}/(M^2_{L,11}-M^2_{L,33})$) instead of 
the SUSY parameters in the sneutrino sector as our input parameters.

In addition to the MSSM paramters listed above the decay
widths of the rare lepton decays depend also 
on other MSSM parameters, which we fix throughout this study.
These are the soft SUSY breaking parameters in the charged slepton sector,
which we take as $M_{E,11}=700$~GeV, $M_{E,22}=800$~GeV, $M_{E,33}=900$~GeV, 
$M_{E,\alpha\neq\beta}=0$, $A_{\alpha\beta}=0$, $\alpha,\beta=1,2,3$,
(for the convention see e.g. \cite{Bartl:2005yy}), 
and the parameter $M_1$ of the neutralino sector, where
we assume the GUT inspired relation 
$|M_1|=(5/3)\tan^2\Theta_W~M_2$, with $M_1<0$.

In Fig.~\ref{fig:fig1}a we show the $\cos2\theta_{13}$ dependence of the
cross section $\sigma(e^+e^-\rightarrow\tilde{\chi}^+_1\tilde{\chi}^-_1)$ 
for three values of $m_{\tilde\nu_3}=(400,600,900)$~GeV with $m_{\tilde\nu_1}=300$~GeV,
$m_{\tilde\nu_2}=350$~GeV, $\mu=1500$~GeV, $M_2=240$~GeV and $\tan\beta=5$.
The resulting chargino masses are $m_{\chi_1}=238$~GeV and $m_{\chi_2}=1505$~GeV.
Fig.~\ref{fig:fig1}b shows the 
appropriate dependence of the branching ratio 
BR($\tau^-\to e^-\gamma$) for the same parameters.
As can be seen in Fig.~\ref{fig:fig1}b, the LFV mixing angle
$\cos2\theta_{13}$ is not restricted and can have any value 
in the range $[-1,1]$. $\cos2\theta_{13}=-1,1$ are the cases 
where lepton flavour is conserved, while
for $\cos2\theta_{13}=0$ LFV is maximal, and the mass eigenstates 
$\tilde\nu_1$ and $\tilde\nu_3$ are mixtures containing an equal 
amount of $\tilde\nu_e$ and $\tilde\nu_\tau$.

Furthermore, we can see in Fig.~\ref{fig:fig1} that even if the present 
bound on the rare decay $\tau^-\to e^-\gamma$ improves by a factor of thousand
the cross section for $e^+e^-\rightarrow\tilde{\chi}^+_1\tilde{\chi}^-_1$ can 
change by a factor two when comparing the cross section for the
lepton flavour conserving (LFC) case $\cos2\theta_{13}=1$ with the one 
for which LFV is maximal ($\cos2\theta_{13}=0$).
We note that the branching ratio BR$(\tau^-\to e^-e^+e^-)$ is 1--2 
orders of magnitude smaller than BR$(\tau^-\to e^-\gamma)$. 
We find that although the size of the cross section strongly depends
on the choice of the beam polarizations, the relative size of 
the cross section with and without LFV is almost independent of it.

\begin{figure}[t]
\setlength{\unitlength}{1mm}
\begin{center}
\begin{picture}(150,50)
\put(-57,-175){\mbox{\epsfig{figure=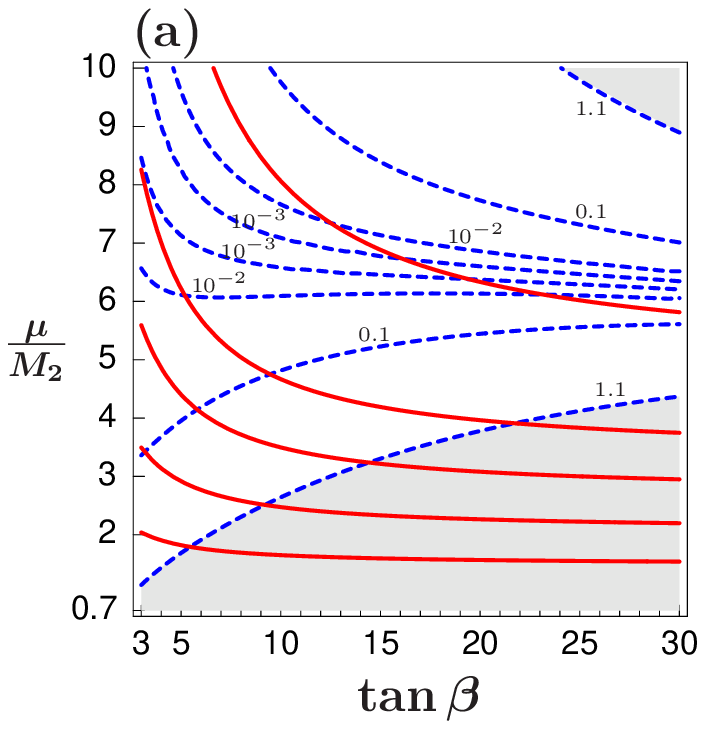,height=25cm,width=23cm}}}
\put(-57,-240){\mbox{\epsfig{figure=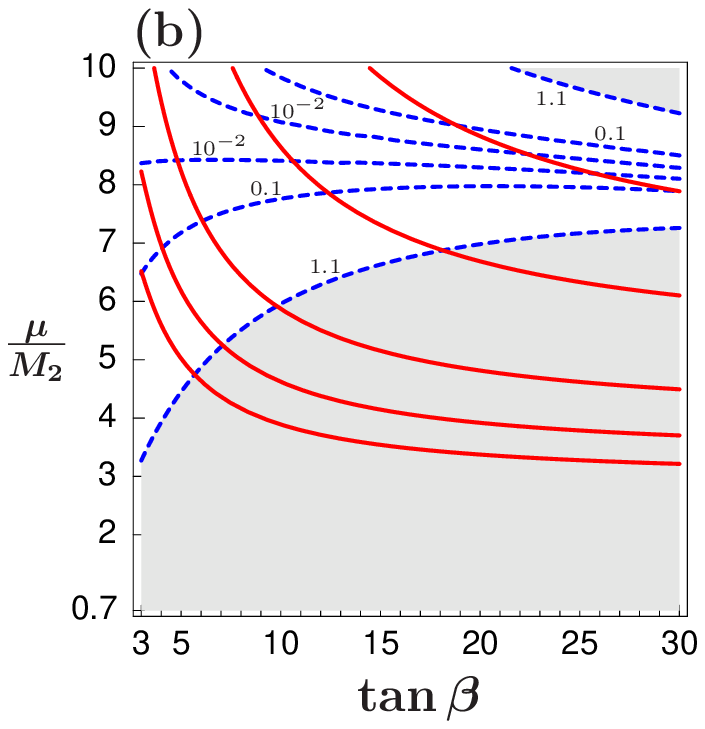,height=25.cm,width=23cm}}}
\end{picture}
\end{center}
\vskip8cm
\caption{Contours of $10^7\cdot$BR($\tau^-\to e^-\gamma$) (dashed lines)
and $\sigma^{\rm LFV}_{11}/\sigma^{\rm LFC}_{11}$ (solid lines) in the
$\mu/M_2$--$\tan\beta$ plane. In (a) we have $m_{\tilde\nu_3}=400$~GeV with the 
contours $\sigma^{\rm LFV}_{11}/\sigma^{\rm LFC}_{11}=(1.5,1.7,1.8,1.85,1.9)$
(bottom-up), and in (b) we have  $m_{\tilde\nu_3}=900$~GeV with the 
contours $\sigma^{\rm LFV}_{11}/\sigma^{\rm LFC}_{11}=
(4,4.1,4.2,4.3,4.35)$ (bottom-up). The shaded areas in (a) and (b)
mark the regions excluded by the present experimental limit 
BR$(\tau^-\to e^-\gamma)<1.1\cdot 10^{-7}$.}
\label{fig:fig2}
\end{figure}

In Fig.~\ref{fig:fig2} we plot the contours of the branching ratio 
$10^7\cdot$BR($\tau^-\to e^-\gamma$) (dashed lines) and the contours of the ratio 
$\sigma^{\rm LFV}_{11}/\sigma^{\rm LFC}_{11}$ (solid lines) in the $\mu/M_2$--$\tan\beta$
plane, where we have used the abbreviations $\sigma^{\rm LFV}_{11}\equiv
\sigma(e^+e^-\rightarrow\tilde{\chi}^+_1\tilde{\chi}^-_1)$ for 
maximal LFV ($\cos2\theta_{13}=0$) and $\sigma^{\rm LFC}_{11}\equiv
\sigma(e^+e^-\rightarrow\tilde{\chi}^+_1\tilde{\chi}^-_1)$ for
the lepton flavour conserving case ($\cos2\theta_{13}=1$).
The other MSSM parameters are the same as in Fig.~\ref{fig:fig1}.
In Fig.~\ref{fig:fig2}a we show the result for $m_{\tilde\nu_3}=400$~GeV
where the contours of $\sigma^{\rm LFV}_{11}/\sigma^{\rm LFC}_{11}$
are 1.5, 1.7, 1.8, 1.85 and 1.9 for increasing $\mu/M_2$.
In Fig.~\ref{fig:fig2}b we have chosen $m_{\tilde\nu_3}=900$~GeV and
the contours for $\sigma^{\rm LFV}_{11}/\sigma^{\rm LFC}_{11}$
in this case are 4, 4.1, 4.2, 4.3 and 4.35 for increasing $\mu/M_2$.
As can be seen in Fig.~\ref{fig:fig2}a and b there is a
region in the $\mu/M_2$--$\tan\beta$ plane where the branching ratio
BR$(\tau^-\to e^-\gamma)$ is two to three orders of magnitude below
its present experimental bound and the values of the 
cross section $\sigma(e^+e^-\rightarrow\tilde{\chi}^+_1\tilde{\chi}^-_1)$ in the LFV case
can be about a factor 2 and 4 larger than in the LFC case.
%In this region $\tilde\chi^+_1$ is gaugino--like and $\tan\beta$
%can have any value in the range shown in Fig.~\ref{fig:fig2}.
%

\subsection{$\tilde\nu_e$--$\tilde\nu_\mu$ mixing case}

\begin{figure}[t]
\setlength{\unitlength}{1mm}
\begin{center}
\begin{picture}(150,50)
\put(-45,-140){\mbox{\epsfig{figure=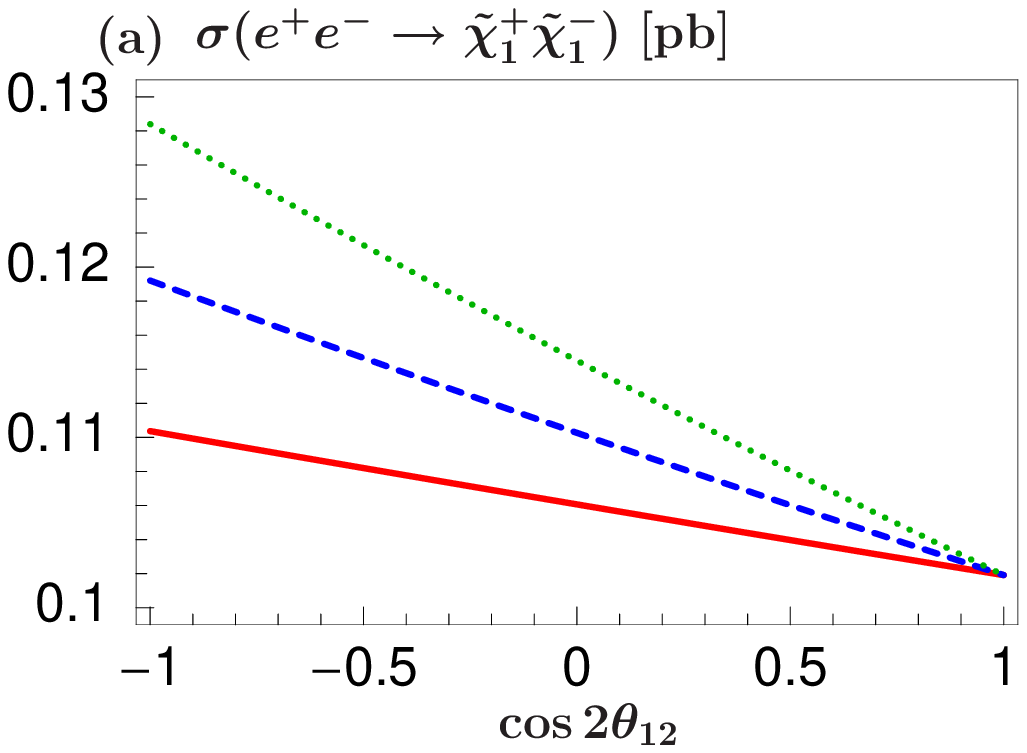,height=21.cm,width=15.4cm}}}
\put(-45,-200){\mbox{\epsfig{figure=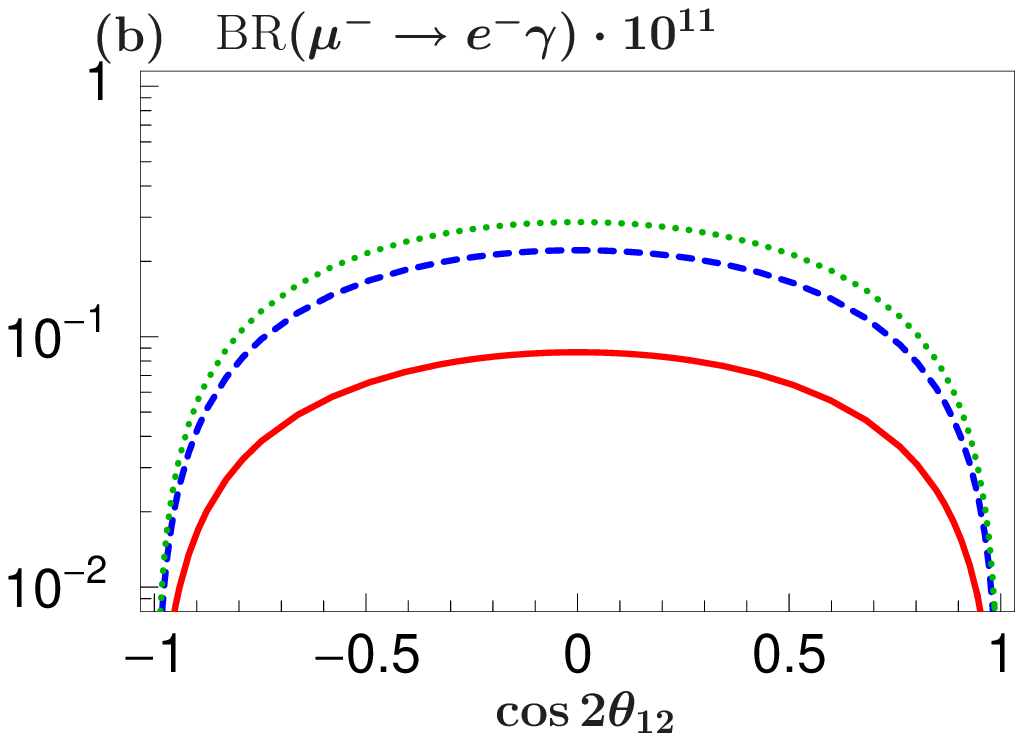,height=21.cm,width=15.4cm}}}
\end{picture}
\end{center}
\vskip6.5cm
\caption{(a) Cross section
$\sigma(e^+e^-\rightarrow\tilde{\chi}^+_1\tilde{\chi}^-_1)$ and
(b) branching ratio BR($\mu^-\to e^-\gamma$) as a function
of $\cos2\theta_{12}$. The three lines correspond to 
$m_{\tilde\nu_2}=305$~GeV (solid line),
310~GeV (dashed line) and 315~GeV (dotted line). 
The other parameters are as specified in the text.}
\label{fig:fig3}
\end{figure}

Now we consider the case of a non-vanishing $M^2_{L,12}$, putting
$M^2_{L,13}$ and $M^2_{L,23}$ to zero.
The size of $M^2_{L,12}$ is strongly restricted by
the experimental upper bounds on the 
LFV processes $\mu^-\to e^-\gamma$ and $\mu^-\to e^- e^+ e^-$
whose sensitivities are about four orders of magnitude larger
than those on LFV tau decays and will improve
substantially in the near future \cite{PSI}.
Similarly as in the previous subsection we take as our input parameters 
the sneutrino masses $m_{\tilde\nu_1}$, $m_{\tilde\nu_2}$, $m_{\tilde\nu_3}$
and the LFV mixing angle $\cos2\theta_{12}$ instead of the soft SUSY breaking parameters 
in the sneutrino mass matrix, Eq.~(\ref{eq:sneutrinomass}).

In Fig.~\ref{fig:fig3}a we show the $\cos2\theta_{12}$ dependence of the
cross section $\sigma(e^+e^-\rightarrow\tilde{\chi}^+_1\tilde{\chi}^-_1)$ 
for three values of $m_{\tilde\nu_2}=(305,310,315)$~GeV with $m_{\tilde\nu_1}=300$~GeV,
$m_{\tilde\nu_3}=500$~GeV, $\mu=1350$~GeV and the other parameters as 
defined in Fig.~\ref{fig:fig1}. 
The chargino masses are $m_{\chi_1}=237$~GeV and $m_{\chi_2}=1355$~GeV.
Fig.~\ref{fig:fig1}b shows the corresponding dependence of the branching ratio 
BR($\mu^-\to e^-\gamma$) for the same parameters.
The LFV mixing angle $\cos2\theta_{12}$ is not restricted
and can have any value in the whole range $[-1,1]$, where
for the values $\cos2\theta_{12}=-1,1$ lepton flavour is
conserved. Once $\cos2\theta_{12}\neq -1,1$
the sneutrinos $\tilde\nu_1$ and $\tilde\nu_2$ are mixtures
of the flavour states $\tilde\nu_e$ and $\tilde\nu_\mu$.
For $\cos2\theta_{12}=0$ they are a mixture containing an equal 
amount of $\tilde\nu_e$ and $\tilde\nu_\mu$, corresponding
to the case of maximal LFV. By comparing the cross sections
of the LFC case with $\cos2\theta_{12}=1$ and the case where 
LFV is maximal ($\cos2\theta_{12}=0$), we see from Fig.~\ref{fig:fig3}a
that the difference can be about 12\%.
We find that the branching ratio BR$(\mu^-\to e^-e^+e^-)$ is 1--2 
orders of magnitude below its present bound in this scenario.

\section{Conclusions \label{sec:3}}

In conclusion, we have pointed out that the influence
of LFV can enormously change the predicted values of the
chargino production cross sections at the ILC.
Studying the production cross section for the reaction
$e^+e^-\to\ti\chi^+_1\ti\chi^-_1$, 
we have shown that it can change by a factor of 2 or more through non--vanishing
LFV parameters which are consistent at the same time with
the present limits on LFV rare lepton decays.
Moreover, we have pointed out that this statement holds
even in the case where the limit on BR($\tau^-\to e^-\gamma$) improves 
by a factor of thousand. In the effort of reconstructing the underlying
model parameters from measurements of chargino production cross sections,
one inevitably has to take the LFV parameters into account. 
This can done by measurements of lepton
flavour violating production and decay rates of SUSY particles 
at the ILC. 
For example, a measurement of the event rates 
for the reaction $e^+e^-\to\tilde\nu\bar{\tilde\nu}\to\tau^+ e^-\tilde\chi^+_1\tilde\chi^-_1$
may allow us to determine the LFV mixing angle $\cos2\theta_{13}$ in the 
sneutrino sector \cite{Nomura:2000zb,SneuLFV}.

\section*{Acknowledgements}

This work is supported by the 'Fonds zur F\"orderung der
wissenschaftlichen Forschung' (FWF) of Austria, project. No. P18959-N16
and by the EU under the MRTN-CT-2006-035505
network programme.

\end{document}